\documentclass[%
 reprint,
 amsmath,amssymb,
 aps,
]{revtex4-2}

\usepackage{amsmath}

\usepackage{xcolor}
\usepackage{graphicx}% Include figure files
\usepackage{dcolumn}% Align table columns on decimal point
\usepackage{bm}% bold math
\begin{document}

\preprint{APS/123-QED}

\title{Assessing High-Order Links in Cardiovascular and Respiratory Networks via Static and
Dynamic Information Measures}


\author{Gorana Mijatovic}
 \altaffiliation{Faculty of Technical Sciences, University of Novi Sad, Serbia}
 \email{gorana86@uns.ac.rs}
 
\author{Laura Sparacino}%
 \altaffiliation{Department of Engineering, University of Palermo, Italy}
 
\author{Yuri Antonacci}
\altaffiliation{Department of Engineering, University of Palermo, Italy}

\author{ Michal Javorka}
\altaffiliation{Department of Physiology and Biomedical Center Martin, Jessenius Faculty of Medicine, Comenius University, Martin,
Slovakia}

\author{Daniele Marinazzo}
\altaffiliation{Department of Data Analysis, University of Ghent, Belgium}

\author{Sebastiano Stramaglia}
\altaffiliation{Department of Physics, University of Bari Aldo Moro, and INFN Sezione di Bari, Italy}

\author{Luca Faes}
\altaffiliation{Department of Engineering, University of Palermo, Italy}
\email{luca.faes@unipa.it}

\date{\today}% It is always \today, today,

\begin{abstract}
The network representation is becoming increasingly popular for the description of cardiovascular interactions based on the analysis of multiple simultaneously collected variables. However, the traditional methods to assess network links based on pairwise interaction measures cannot reveal high-order effects involving more than two nodes, and are not appropriate to infer the underlying network topology. To address these limitations, here we introduce a framework which combines the assessment of high-order interactions with statistical inference for the characterization of the functional links sustaining physiological networks. 
The framework develops information-theoretic measures quantifying how two nodes interact in a redundant or synergistic way with the rest of the network, and employs these measures for reconstructing the functional structure of the network.
The measures are implemented for both static and dynamic networks mapped respectively by random variables and random processes using plug-in and model-based entropy estimators. 
The validation on theoretical and numerical simulated networks documents the ability of the framework to represent high-order interactions as networks and to detect statistical structures associated to cascade, common drive and common target effects. The application to cardiovascular networks mapped by the beat-to-beat variability of heart rate, respiration, arterial pressure, cardiac output and vascular resistance allowed noninvasive characterization of several mechanisms of \textcolor {black} {cardiovascular} control operating in resting state and during orthostatic stress. 
Our approach brings to new comprehensive assessment of physiological interactions and complements existing strategies for the classification of \textcolor {black} {pathophysiological} states.
\end{abstract}

%\keywords{cardiovascular variability, higher-order interactions, information theory, network physiology, redundancy and synergy}%Use showkeys class option if keyword
\maketitle

\section{INTRODUCTION}
In recent years, there has been a growing focus on exploring interactions within the cardiovascular and respiratory systems aimed at gaining insights into diverse \textcolor {black} {pathophysiological}  mechanisms \cite{schulz2013cardiovascular}. This surge of interest is driven by fast advancements in the techniques for recording and processing biomedical signals, which nowadays offer unprecedented opportunities for the simultaneous monitoring of multiple cardiovascular parameters. In turn, biomedical signal processing methods are shifting towards the multivariate analysis of several parameters, which naturally leads to representing physiological systems as interconnected networks whose nodes and links are associated respectively to the parameters and to the interactions between them. This representation is very popular in the field of Network Physiology \cite{ivanov2021new} where cardiovascular, cardiorespiratory interactions and others are studied in a wide range of healthy and diseased states \cite{cohen2002short, faes2013mechanisms,elstad2018cardiorespiratory,santiago2022effects}.

The analysis of physiological networks typically relies on pairwise measures, i.e. measures that quantify the existence and strength of a link focusing exclusively on the activity of the two nodes connected by it. Pairwise measures such as the linear correlation, the spectral coherence or the mutual information (MI) and MI rate (MIR) have been successfully applied to assess link strength in cardiovascular and respiratory networks \cite{schulz2013cardiovascular,cohen2002short, porta2002quantifying, hoyer2002mutual, mijatovic2022measuring}. Nevertheless, these measures suffer from two main limitations: they are not appropriate to infer the topological structure of the analyzed functional network, because they do not allow to distinguish direct relations between two nodes from indirect relations caused by common drive or cascade effects \cite{sanchez2021combining, reid2019advancing}; and they cannot assess functional high-order interactions (HOIs), i.e. interactions that result from the combined activity of more than two nodes and encompass the statistical concepts of synergy and redundancy \cite{rosas2019quantifying,battiston2020networks}. 
To address these limitations, several approaches have been proposed which separately consider static or dynamic network systems, i.e. systems where the temporal correlations within and between the node activities are neglected or investigated. In particular, the inference of network structure has been faced combining correlation and partial correlation \cite{reid2019advancing} for static systems, and exploiting the concept of multivariate Granger causality \cite{barrett2010multivariate} for dynamic systems. The analysis of HOIs is performed via emerging information-theoretic methods like the partial information decomposition \cite{williams2010nonnegative} or the so-called O-information \cite{rosas2019quantifying}, originally proposed for random variables and recently extended to dynamic systems \cite{krohova2019multiscale,faes2022new}.
Interestingly, while  network reconstruction and HOIs assessment are both grounded on multivariate analysis, their interrelationships remain unexplored, and a unified approach for examining HOIs at the level of the network links which can be exploited also for structural inference is currently absent.

The present study introduces a comprehensive approach that combines the assessment of HOIs with statistical inference for the characterization of the functional links in complex physiological networks. The approach is designed for generic network systems by using information-theoretic concepts, and is then contextualized to static systems mapped by random variables and dynamic systems mapped by random processes, respectively using measures of entropy and entropy rate. Data-efficient estimators are implemented for the proposed interaction measures, which are thoroughly validated on simulated network systems. 
Then, the framework is applied to cardiovascular networks identified measuring several physiological parameters reflecting the heart period or its variation, the respiratory amplitude or phase, the systolic and diastolic pressure, the cardiac output and the vascular resistance. Applications are focused on showing the usefulness of analyzing HOIs at the level of the network links to characterize both well-established and less explored cardiovascular regulatory mechanisms in different physiological conditions.

\section {Methods}

In this work, given a network system composed of several possibly interconnected units, we consider the problem of quantifying the link between the units in a probabilistic framework. Sect. II.A introduces the measures used to investigate the interaction between any two system units while taking into account the rest of the network, provides an information-theoretic explanation of these measures and illustrates their use for network inference. Sect. II.B describes how these measures can be formalized in terms of the entropy of random variables in the case of static systems observed regardless of the flow of time, or in terms of the entropy rate of random processes in the case of dynamic systems evolving in time. Sect. II.C describes the practical computation based on plug-in estimators for discrete random variables and parametric estimators for continuous random processes.

\subsection{Measuring Interactions in Network Systems}
Let us consider an observer measuring a system composed by $M$ units, $\mathcal{S}=\{\mathcal{S}_1,\ldots,\mathcal{S}_M\}$, and focus on the two units 
$\mathcal{X}=\mathcal{S}_i$ and $\mathcal{Y}=\mathcal{S}_j$ while collecting the remaining $M-2$ units in the group $\mathcal{Z}=\mathcal{S}\setminus{\{\mathcal{X},\mathcal{Y}\}}$.
In the framework of information theory, the interaction between  $\mathcal{X}$ and $\mathcal{Y}$ is assessed quantifying the information shared (IS) between them, or the conditional information shared (cIS) between them but not with $\mathcal{Z}$; while these quantities are kept generic for now, in the next section they will be expressed by using mutual information or mutual information rate, respectively when contextualized to static or dynamic systems.

For concreteness, let us denote with $\mathcal{H_X}$ and $\mathcal{H_Y}$ the information contained in the units $\mathcal{X}$ and $\mathcal{Y}$, and with $\mathcal{H_{X,Y}}$ the joint information contained in the two units taken together. Then, the IS between $\mathcal{X}$ and $\mathcal{Y}$ is given by
\begin{equation}
\mathcal{I_{X;Y}}=\mathcal{H_X}+\mathcal{H_Y}-\mathcal{H_{X,Y}},
   \label{SharedInfo}
\end{equation}
while the cIS between $\mathcal{X}$ and $\mathcal{Y}$ given  $\mathcal{Z}$ is given by
\begin{equation}
\mathcal{I_{X;Y|Z}}=\mathcal{I_{X;Y,Z}}-\mathcal{I_{X;Z}}.
   \label{CondSharedInfo}
\end{equation}

The measures in (\ref{SharedInfo}) and (\ref{CondSharedInfo}) quantify the link between the two analyzed units from a bivariate or multivariate perspective, and are illustrated by using Venn diagrams in Fig. 1a.
Importantly, the comparison between these two quantities highlights the balance between the statistical concepts of redundancy and synergy in the observed network system. In particular, %we define the difference between (\ref{SharedInfo}) and (\ref{CondSharedInfo}) as 
the net information shared (nIS) between $\{ \mathcal{X},\mathcal{Y} \}$ and $\mathcal{Z}$, defined as
\begin{equation}
\mathcal{I_{X;Y;Z}}=\mathcal{I_{X;Y}}-\mathcal{I_{X;Y|Z}}.
   \label{InteractionInfo}
\end{equation}
quantifies the interaction between the two analyzed units and the rest of the system, and can be either positive or negative denoting respectively the prevalence of redundancy or synergy.
%This measure can assume both positive and negative values, denoting respectively the prevalence of redundancy and the prevalence of synergy in the interaction between the two analyzed units and the rest of the system.
Specifically, when $\mathcal{I_{X;Y;Z}}>0$, the knowledge of $\mathcal{Z}$ reduces the information shared by $\mathcal{X}$ and $\mathcal{Y}$, thus indicating that (part of) the statistical dependence between $\mathcal{X}$ and $\mathcal{Y}$ is suppressed when $\mathcal{Z}$ is observed.
%$\mathcal{Z}$ explains part of the interaction between $\mathcal{X}$ and $\mathcal{Y}$;
On the contrary, when $\mathcal{I_{X;Y;Z}}<0$, the knowledge of $\mathcal{Z}$ increases the information shared by $\mathcal{X}$ and $\mathcal{Y}$, thus indicating that (part of) the statistical dependence between $\mathcal{X}$ and $\mathcal{Y}$ emerges when $\mathcal{Z}$ is observed.

To emphasize the balance between redundancy and synergy in the interaction among the two observed units $\mathcal{X}$ and $\mathcal{Y}$ and the rest of the system $\mathcal{Z}$, and to retrieve information about the network topology from such interaction, we define the so-called \textit{B-index} (shorthand for redundancy/synergy balance) by normalizing the nIS as follows: 
\begin{equation}
\mathcal{B_{X;Y}}=\dfrac{\mathcal{I_{X;Y;Z}}}{\max\{ \mathcal{I_{X;Y}} , \mathcal{I_{X;Y|Z}}\}}.
\label{B-Index}
\end{equation}
In (\ref{B-Index}), the B-index is computed dividing the difference between $\mathcal{I_{X;Y}}$ and $\mathcal{I_{X;Y|Z}}$ to their maximum, so as to obtain a measure ranging between -1 and 1 (see Fig. 1b). The limit values highlight a full imbalance between redundancy and synergy that relates to specific network topologies: $\mathcal{B_{X;Y}}=1$ corresponds to maximum redundancy, occurring when the interaction between $\mathcal{X}$ and $\mathcal{Y}$ is fully explained by the rest of the network and reflecting a relation of common driver 
%($\mathcal{X}\leftarrow \mathcal{Z} \rightarrow \mathcal{Y}$) or cascade ($\mathcal{X} \rightarrow \mathcal{Z} \rightarrow \mathcal{Y}$ or $\mathcal{Y} \rightarrow \mathcal{Z} \rightarrow \mathcal{X}$) 
or cascade (Fig. 1b, case 1); $\mathcal{B_{X;Y}}=-1$ corresponds to maximum synergy, occurring when the interaction between $\mathcal{X}$ and $\mathcal{Y}$ arises fully from their effect on $\mathcal{Z}$  and reflecting a common target relation %($\mathcal{X}\rightarrow \mathcal{Z} \leftarrow \mathcal{Y}$, 
(Fig. 1b, case 2). In both cases, the nodes mapped by $\mathcal{X}$ and $\mathcal{Y}$ are topologically disconnected. The two nodes are disconnected also when both $\mathcal{I_{X;Y}}$ and $\mathcal{I_{X;Y|Z}}$ are null, resulting in non-defined B-index and describing a situation in which at least one between $\mathcal{X}$ and $\mathcal{Y}$ is isolated from the rest of the network (Fig. 1b, case 3).
Intermediate values of the B-index ($-1<\mathcal{B_{X;Y}}<1$) are obtained when both $\mathcal{I_{X;Y}}$ and $\mathcal{I_{X;Y|Z}}$ are non-null (Fig. 1b, case 4). In this situation, larger values of $\mathcal{I_{X;Y}}$ denote prevalence of redundancy, while  larger values of $\mathcal{I_{X;Y|Z}}$ denote prevalence of synergy, which topologically corresponds to many possible configurations; the case of identical non-null values denotes perfect balance between synergy and redundancy, indicating that the two nodes are linked to each other but disconnected from the rest of the network.

\begin{figure*} [tbp]
\centering
\includegraphics[scale=0.9]{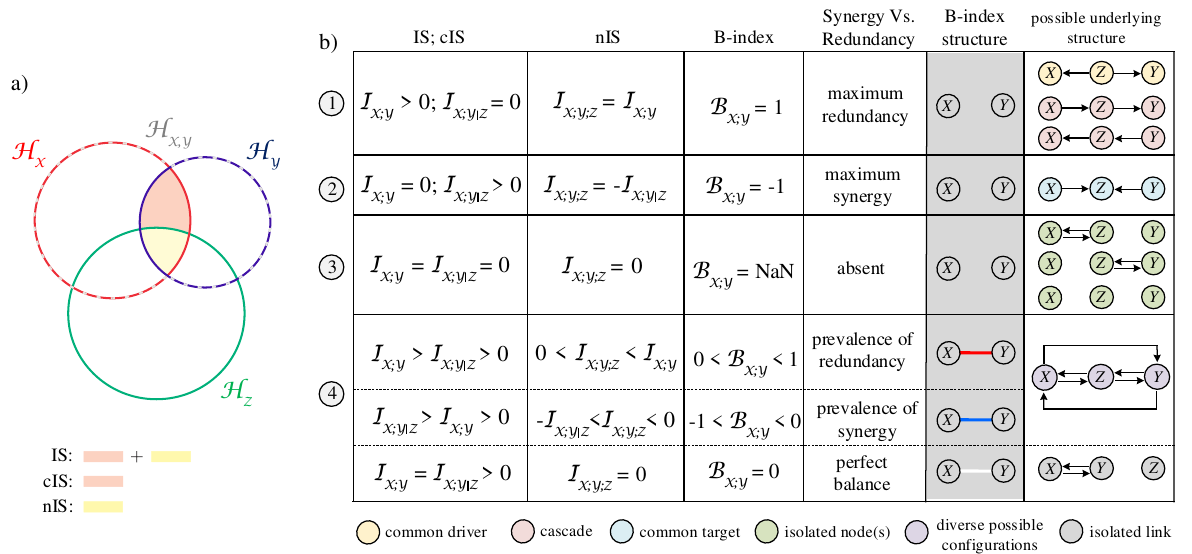}
\caption{Illustration of the measures proposed to assess high-order links between two units $\mathcal{X}$ and $\mathcal{Y}$ of a network system. (a) Venn diagram explaining the derivation of the measures of information shared between $\mathcal{X}$ and $\mathcal{Y}$ (IS, $\mathcal{I}_{\mathcal{X};\mathcal{Y}}$), conditional IS between $\mathcal{X}$ and $\mathcal{Y}$ given the rest of the network $\mathcal{Z}$ (cIS, $\mathcal{I}_{\mathcal{X};\mathcal{Y}|\mathcal{Z}}$), and the net information shared (nIS, $\mathcal{I}_{\mathcal{X};\mathcal{Y};\mathcal{Z}}$), from the information contained in the units ($\mathcal{H_X}$, $\mathcal{H_Y}$, $\mathcal{H_Z}$, information of $\mathcal{X}$, $\mathcal{Y}$, $\mathcal{Z}$; $\mathcal{H_{X,Y}}$, joint information of  $\mathcal{X}$ and $\mathcal{Y}$).
b) Classification of the redundant/synergistic nature of the interactions and of the types of possible underlying structural mechanisms, based on the assessment of zero and strictly positive values of IS and cIS leading to characteristic values of the B-index.}
\label{00_B}
\end{figure*}

The configurations described above will be illustrated in the theoretical and simulated examples of Sect. III.
Note that, in practical analysis, the presence of strictly positive or null values of $\mathcal{I_{X;Y}}$ and $\mathcal{I_{X;Y|Z}}$ leading to the limit values of the B-index, is tested by surrogate data approaches (Sect. II.C).

\subsection{Implementation for Static and Dynamic Systems}
In this section, the concept of information shared between the units of a generic network system is contextualized to make it usable in real-world applications. Specifically, if the quantities used to observe the system states do not depend on time (e.g., they are indexes measured across subjects), the system is denoted as \textit{static} and is described in terms of random variables. On the contrary, a \textit{dynamic} system evolves in time and, as such, is described by random processes, i.e. collections of random variables sorted in temporal order.

In practice, a static system $\mathcal{S}$ composed by $M$ units is described by the $M-$dimensional random variable $S=[S_1 \cdots S_M]$, and the states assumed by the units $\mathcal{X}$, $\mathcal{Y}$ and $\mathcal{Z}$ are mapped by the random variables $X$, $Y$ and $Z$.
%where $S_i$ is the random variable mapping the state of the $i^{th}$ system unit $\mathcal{S}_i, i=1,\ldots,M$.
In this case, the information contained in the unit $\mathcal{X}$ is quantified by the entropy of the variable $X$, $H(X)=-\mathbb{E}[\log p(x)]$, where $p(x)$ is the probability density of $X$ and $\mathbb{E}[\cdot]$ is the expectation operator (the same holds for $Y$ and $Z$). Then, the information shared between the variables $X$ and $Y$ is quantified by the mutual information (MI), which is computed by using the entropies in (\ref{SharedInfo}) for IS to get $I(X;Y)=H(X)+H(Y)-H(X,Y)$,
while the cIS between $X$ and $Y$ given $Z$ is quantified by the conditional MI (cMI) obtained as $I(X;Y|Z)=I(X;Y,Z)-I(X;Z)$ \cite{cover1991elements}. The MI and cMI are then readily inserted in (\ref{InteractionInfo}) and (\ref{B-Index}) to obtain the nIS  and the B-index relevant to the analyzed network of random variables. Note that, in this setting where it is computed for random variables, the nIS corresponds to the well known interaction information \cite{mcgill1954multivariate}; this measure, used as in our case to describe how two units of a network system interact with the remaining units, is referred to as the local O-Information \cite{rosas2019quantifying}.

When the analyzed system $\mathcal{S}$ is dynamic, it is more properly described by a vector random process $S_n=[S_{1,n},\ldots,S_{M,n}]$, where the temporal dependence is denoted by the time counter $n$ and $S_{i,n}$ denotes the variable sampling the $i^{th}$ process at the $n^{th}$ time step ($i=1,\ldots,M; n \in \mathbb{Z}$ for discrete-time processes); the equivalent notation evidencing the dynamics of the units $\{\mathcal{X},\mathcal{Y},\mathcal{Z}\}$ is $S_n=[X_n,Y_n,Z_n]$. In this case, the information-theoretic measure typically used to analyze the temporal evolution of the unit $\mathcal{X}$ is the entropy rate,  defined for a stationary process using the conditional entropy $H_X=H(X_n|X_{n-1},X_{n-2},\ldots)$. The entropy rate quantifies the rate of generation of new information in the process $X$; when combined with the entropy rate of $Y$, $H_Y$, and with the joint entropy rate of $X$ and $Y$, $H_{X,Y}$, it reveals the IS of the two processes measured by the so-called mutual information rate (MIR), $I_{X;Y}=H_{X}+H_{Y}-H_{X,Y}$ \cite{duncan1970}. Based on these definitions, in a network of random processes the cIS becomes the conditional MIR (cMIR) $I_{X;Y|Z}$ obtained using $I_{X;Y,Z}$ and $I_{X;Z}$ in (\ref{CondSharedInfo}), the nIS becomes the interaction information rate $I_{X;Y;Z}$ obtained using $I_{X;Y}$ and $I_{X;Y|Z}$ in (\ref{InteractionInfo}), and the B-index becomes a "B-index rate" $B_{X;Y}$ obtained using $I_{X;Y;Z}$, $I_{X;Y}$ and $I_{X;Y|Z}$ in (\ref{B-Index}).

\subsection{Practical Computation}

This section describes how the interaction measures defined in Sect. II.A and particularized to static or dynamic systems in Sect. II.B can be computed in practice from observations of the states assumed by the units composing the observed network system. In particular, we show how the static measures based on MI can be obtained from observations of discrete random variables collected as sequences of symbols using plug-in entropy estimators, and how the dynamic measures based on MIR can be obtained from realizations of continuous random processes collected as time series using parametric estimators. 

\subsubsection{Discrete Random Variables}

When $X$, $Y$, and each component of the vector $Z$ are discrete random variables taking values in the sets $\mathcal{A}_X,\mathcal{A}_Y$, and $\mathcal{A}_Z$, formed respectively by $Q_X$, $Q_Y$ and $Q_Z$ symbols, the computation of MI and cMI follows the well-known formulations \cite{cover1991elements}:
\begin{subequations}
   \begin{align}
I(X;Y) = \sum\limits_{x \in \mathcal{A}_X, y \in \mathcal{A}_Y}  p(x,y) \log\frac{p(x,y)}{p(x)p(y)}, \\
I(X;Y|Z) = \sum\limits_{x \in \mathcal{A}_X, y \in \mathcal{A}_Y, z \in \mathcal{A}_Z} p(x,y,z) \log\frac{p(x,y|\textbf{\textit{z}})}{p(x|\textbf{\textit{z}})p(y|\textbf{\textit{z}})},
\end{align}
\label{MI_CMI_d}
\end{subequations}
where $p(\cdot)$,  $p(\cdot, \cdot)$, and $p(\cdot|\cdot)$ denote the marginal, joint, and conditional probability density of the analyzed variables. Given the MI and cMI measures in (\ref{MI_CMI_d}), the interaction information and the B-index are obtained using (\ref{InteractionInfo}) and (\ref{B-Index}).

In practical analysis, the network measures are computed from observations of the random variables $X$, $Y$ and $Z$ available in the form of synchronous sequences of symbols first estimating the probabilities as the frequency of occurrence of the relevant combination of symbols within the sequences, and then plugging the probability estimates into (\ref{MI_CMI_d}) to obtain estimates of MI and cMI.

\subsubsection{Continuous Random Processes}
When the system units are represented by a vector random process $S=\{X,Y,Z\}$, a common approach to describe the process dynamics is to use the linear vector autoregressive (VAR) model \cite{lutkepohl2005new}
\begin{equation}
S_n=\sum_{k=1}^{p}\textbf{A}_kS_{n-k}+U_n,
\label{fullVAR}
\end{equation}
where $S_n$ and $S_{n-k}$ are the $M-$dimensional variables sampling the process at the present time step $n$ and at $k$ steps in the past, $\textbf{A}_k$ is an $M \times M$ coefficient matrix and $U$ is an i.i.d. innovation process with $M \times M$ covariance matrix $\mathbf{\Sigma}_U$.

Given the full-model (\ref{fullVAR}), it is possible to define  restricted models which describe the dynamics of subsets of processes. In particular, the individual dynamics of the processes $X$ and $Y$, and their joint dynamics described by the bivariate process $W=\{X,Y\}$, are captured by the restricted models
\begin{subequations}
   \begin{align}
X_n=\sum_{k=1}^{q}A^{(x)}_kX_{n-k}+U_{x,n}, \\
Y_n=\sum_{k=1}^{q}A^{(y)}_kY_{n-k}+U_{y,n}, \\
W_n=\sum_{k=1}^{q}\textbf{A}^{(w)}_kW_{n-k}+U_{w,n},
\end{align}
\label{restrictedVAR}
\end{subequations}
where $q$ is the order of the restricted model (theoretically $q \rightarrow \infty$). The parameters of the restricted models, i.e. the coefficients $A^{(x)}_k$, $A^{(y)}_k$, $\textbf{A}^{(w)}_k$, and the covariance of the residuals $\Sigma_{U_x}$, $\Sigma_{U_y}$, $\mathbf{\Sigma}_W$, can be derived from the parameters of the full model (\ref{fullVAR}) $\textbf{A}_k$ and $\mathbf{\Sigma}_U$, through a procedure that solves the Yule-Walker equations to derive the covariance structure of $S$ and then reorganizes such structure to relate it to the covariances of $X$, $Y$ or $W$ (the procedure is described in detail in Ref. \cite{faes2017information}).
Then, under the assumption of joint Gaussianity for the overall process $S$, the information measures capturing the network interactions can be derived straightly from the covariances of the residuals of the restricted models. Specifically, the entropy rates of $X$, $Y$ and $W$ are obtained as
\begin{subequations}
   \begin{align}
H_X=\frac{1}{2} \log 2\pi e \Sigma_{U_x}, \\
H_Y=\frac{1}{2} \log 2\pi e \Sigma_{U_y}, \\
H_{X,Y}=\frac{1}{2} \log (2\pi e)^2 \mathbf{\Sigma}_{U_w},
\end{align}
\label{EntropyRates}
\end{subequations}
from which the MIR is computed as
\begin{equation}
I_{X;Y}=H_X+H_Y-H_{W}=\frac{1}{2} \log \frac{\Sigma_{U_x} \Sigma_{U_y}}{\mathbf{\Sigma}_{U_w}}.
\label{MIRgauss}
\end{equation}

The procedure described above can be repeated to define restricted models capturing the dynamics of the vector process $Z$ as well as of the joint processes $V=\{X,Z\}$ and $R=\{Y,Z\}$ using VAR formulations as in (\ref{restrictedVAR}), and then to compute the entropy rates of $Z$, $V$ and $R$ as in (\ref{EntropyRates}). This allows to obtain formulations of the MIR terms $I_{X;Z}$ and $I_{X;Y,Z}=I_{X;R}$
\begin{subequations}
   \begin{align}
I_{X;Z}=H_X+H_Z-H_{V}=\frac{1}{2} \log \frac{\Sigma_{U_x} {\mathbf{\Sigma}_{U_z}}}{\mathbf{\Sigma}_{U_v}}, \\
I_{X;Y,Z}=H_X+H_R-H_{S}=\frac{1}{2} \log \frac{\Sigma_{U_x} {\mathbf{\Sigma}_{U_r}}}{\mathbf{\Sigma}_{U}},
\end{align}
\label{MIRgaussZ}
\end{subequations}
\noindent{from which the cIS $I_{X;Y|Z}$ is computed as in (\ref{CondSharedInfo}) subtracting (\ref{MIRgaussZ}a) from (\ref{MIRgaussZ}b), and the nIS $I_{X;Y;Z}$  and B-index $B_{X;Y}$  are computed as in (\ref{InteractionInfo}) and (\ref{B-Index}).}

In practical analysis, the network measures are computed from realizations of the random processes $X$, $Y$ and $Z$ available in the form of synchronous time series estimating the parameters of the full model (\ref{fullVAR}) by means of least squares identification \cite{lutkepohl2005new}, deriving the covariances of the residuals of the restricted models by the procedure described in \cite{faes2017information}, and then plugging these covariances into (\ref{MIRgauss}) and (\ref{MIRgaussZ}) to obtain estimates of the MIR and the conditional MIR. The order $p$ of the full model is typically estimated using information-theoretic criteria, while the order $q$ of the restricted models is set at high values to capture the decay of the correlations at increasing lags (in this study, $p$ is optimized between 1 and 20 using the Akaike criterion \cite{akaike1974new} and $q$ is set to 20 \cite{faes2017information}).

\subsubsection{Statistical Significance}
An important issue in practical analysis, which has relevance for the identification of limit values of the B-index and for the corresponding inference of the network structure, is the assessment of the statistical significance of the IS and cIS measures. Such assessment was performed in this work using the method of surrogate data \cite{schreiber2000surrogate}. This method is based on generating, from the measured data relevant to the analyzed network $\mathcal{S}=\{ \mathcal{X}, \mathcal{Y}, \mathcal{Z} \}$, sets of surrogate data which mimic the individual properties of the units $\mathcal{X}$ and $\mathcal{Y}$, but destroy the interaction between them. Under the null hypothesis of absence of IS ($\mathcal{I_{X;Y}}=0$) or absence of cIS ($\mathcal{I_{X;Y|Z}}=0$), the values of IS and cIS computed from the original data are compared with the distribution of IS/cIS computed from the surrogate sets using a test based on percentiles, run with significance $\alpha$ (in this work, $\alpha=0.05$); then, the null hypothesis is rejected (accepted) and the original IS/cIS measure is deemed as statistically significant (non-significant) if the original values of IS or cIS are larger (smaller) than the $(1-\alpha)^{th}$ percentile of the corresponding surrogate distribution. Accordingly, when the estimates of IS and/or cIS are detected as non significant the B-index is set to 1, $-$1, or to NaN according to Fig. 1b.

In the analysis of static systems mapped by random variables, the procedure described above was implemented using the MI or the cMI as discriminating statistic for the quantification of IS or cIS. In this case, realizations of the multiple observed random variables are available in the form of numeric sequences, and surrogate data are generated by independently shuffling in random order the sequences relevant to each variable, so as to make the surrogate variables independent while preserving their marginal distributions. 
On the other hand, in the analysis of dynamic systems mapped by random processes whose realizations are multivariate time series, the MIR or the cMIR were used as discriminating statistic and surrogate time series were generated using the iterative amplitude-adjusted Fourier Transform (iAAFT) procedure \cite{schreiber2000surrogate} which preserves the individual properties of each time series while destroying the interactions among them.

\section{Validation}

In this Section we illustrate the proposed framework for assessing network interactions in simulations of static systems mapped by binary variables and dynamic systems mapped by continuous processes. The measures presented in Sect. II.A are first described in a theoretical example for which their values can be computed analytically, and then validated on finite-length realizations of simulated network activities. 

\subsection{Theoretical Example}

\begin{figure*} [tbp]
\centering
\includegraphics[scale=0.82]{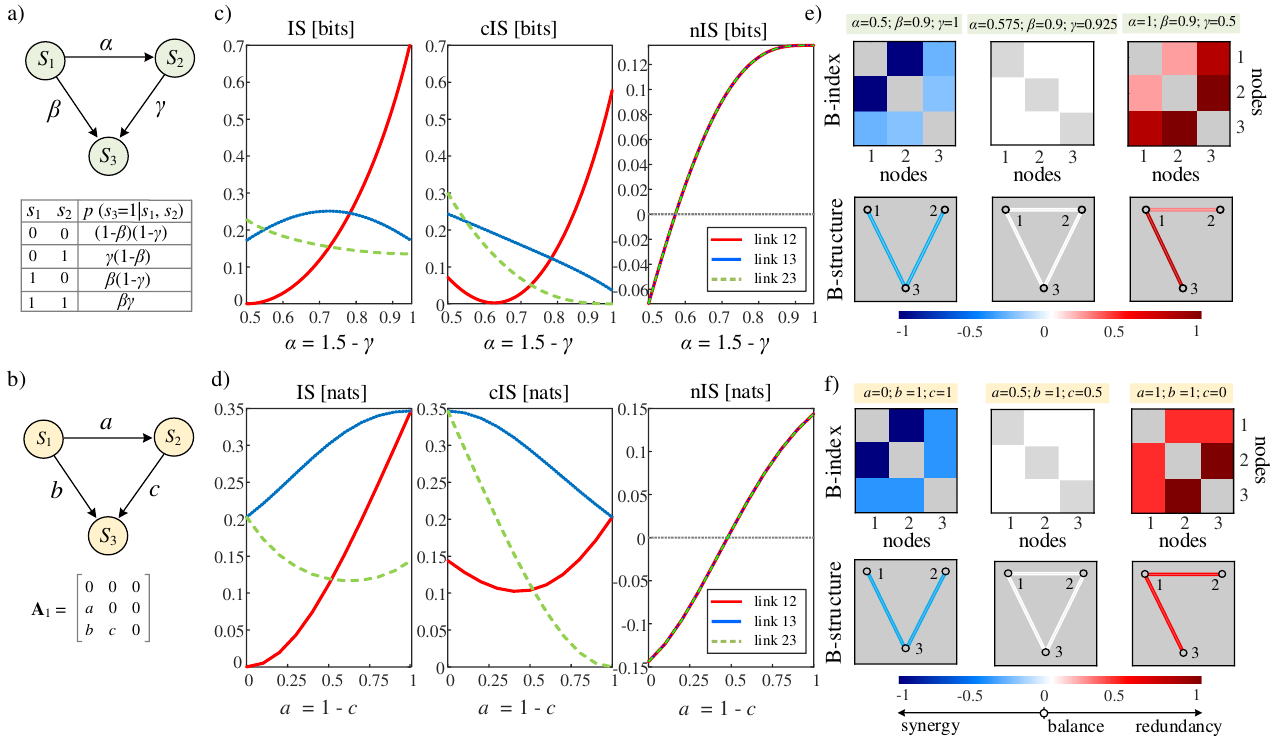}
\caption{Theoretical analysis of a simulated network system with three units mapped by binary random variables (a, static system) or by continuous random processes (b, dynamic system); the strength of the network links is modulated by the probabilities $\alpha, \beta, \gamma$ in (a), and by the VAR model coefficients $a,b,c$ in (b).
The metrics of information shared (IS), conditional information shared (cIS), and net information shared (nIS) are plotted for each network link at varying $\alpha = 1.5 - \gamma$ for the static system (c) or varying $a=1-c$ for the dynamic system (d); this simulates the transition from common target to common driver conditions, mirrored by a shift from synergistic to redundant interaction for both systems. The matrix of normalized B-index values and the corresponding network structure obtained after pruning non-significant links are shown for three representative coupling conditions obtained setting fixed values of $\alpha$, $\beta$, and $\gamma$ (e) or $a$, $b$, and $c$ (f): common target effects reflect pure synergy (left), a fully connected network highlights more balanced synergy/redundancy effects (middle), and  common driver effects reflect pure redundancy (right).}
\label{01_sim_t}
\end{figure*}

We consider the simple case of a system with three units whose interactions are imposed, so as to allow computation of the exact theoretical values of the IS and cIS measures. First, a static system is simulated defining three binary random variables $S_1, S_2, S_3$, which interact as depicted in Fig. \ref{01_sim_t}a: $S_1$ is a binary variable with equiprobable symbols, influencing the state of $S_2$ with probability $\alpha$ ($p(\{s_2=s_1\})=\alpha)$, while both $S_1$ and $S_2$ influence the state of $S_3$ with probabilities $\beta$ and $\gamma$, resulting in the conditional probabilities indicated in Fig. \ref{01_sim_t}a.
Then, a dynamic system with the same causal connections is simulated (Fig. \ref{01_sim_t}b) defining a trivariate VAR process fed by independent Gaussian innovations as in (\ref{fullVAR}), where $M=3, p=1$ and the time-lagged effects are determined by the coefficient matrix $\textbf{\text{A}}_1$.

The analysis was conducted first computing IS, cIS, and nIS between each pair of network nodes at increasing the coupling $S_1 \rightarrow S_2$ and simultaneously decreasing the coupling $S_2 \rightarrow S_3$, with stable coupling $S_1 \rightarrow S_3$; this was achieved by setting $\alpha \in [0.5,1]$, $\gamma=1.5-\alpha$ and $\beta=0.9$ for the static system, and $a \in [0,1]$, $c=1-a$ and $b=1$ for the dynamic system. The results in Fig. \ref{01_sim_t}c,d evidence an increase of IS between $S_1$ and $S_2$ from 0 to high values, a decrease towards zero of cIS between  $S_2$ and $S_3$ given $S_1$, and a shift of nIS from negative to positive values; note that, in these networks with only three nodes, the synergy/redundancy balance nIS is the same for all links.
%. Note that the measure nIS, reflecting the synergy/redundancy balance, is the same for all links in these networks composed by only three nodes.
To further illustrate the shift from synergy to redundancy, the B-index and the corresponding network structure inferred after pruning the links for which both IS and cIS are null were calculated for three representative sets of parameters, reported in Fig. \ref{01_sim_t}e,f. These conditions correspond to the existence of unique common target effects ($\alpha=0.5$ or $a=0$), complete synergy/redundancy balance ($\alpha=0.575$ or $a=0.5$), or unique common drive effects ($\gamma=0.5$ or $c=0$). As shown in Fig. \ref{01_sim_t}e,f, these three conditions are characterized respectively  by highly synergistic interactions (blue colors in the B-index matrix) with absence of the link between $S_1$ and $S_2$, null B-index (white colors) with fully connected network, and highly redundant interactions (red colors) with absence of the link between $S_2$ and $S_3$. Thus, these results illustrate how the proposed framework can detect the structure and the nature of high-order links in simple static or dynamic networks.

\subsection{Simulated data}

In the following, we present the validation of the proposed framework in simulations of static and dynamic systems studied by generating the output data of each network node and estimating the network measures as described in Sect. II.C.
Analyses were performed iterating each simulation 100 times and generating datasets of $N \in \{250, 500, 1000\}$ observations, from which the estimator performance was assessed in terms of sensitivity and specificity of the reconstruction of the known network topology.

\subsubsection{Binary variables}

\begin{figure} [h]
\centering
\includegraphics[scale=1.12]{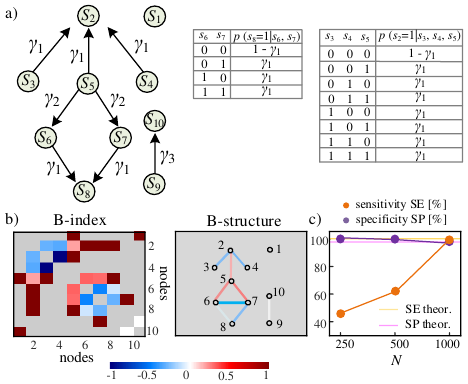}
\caption{Experimental analysis of a simulated static system with ten units mapped by binary variables interconnected via the graph and probability rules shown in (a).
The matrix of B-index values computed from  $N=1000$ observations of the variables and the corresponding connectivity graph (b) evidence the synergistic/redundant nature of the links as well as the proper reconstruction of the network structure, with correct pruning of absent links (isolated nodes/links: $B=$ NaN, grey; common drive, cascade: $B=1$, dark red; common target: $B=-1$, dark blue), except for the false positive detection of a connection between $S_6$ and $S_7$. The performance of network reconstruction, assessed over 100 simulation runs at varying the data length $N$ (c), reveals that sensitivity depends on $N$, while specificity is high but affected by the single false-positive detection for any data length.}
\label{sim_bin}
\end{figure}

The first simulation reproduces a network with $M=10$ nodes connected as depicted in Fig. \ref{sim_bin}a. The node activities are mapped by binary random variables interconnected as follows: $S_1$, $S_3$, $S_4$, $S_5$ and $S_9$ are i.i.d. binary variables with equiprobable symbols; $S_{10}$ is a noisy copy of $S_9$ with coupling strength $\gamma_3=0.8$, (i.e., $p(\{ s_{10}=s_{9} \})=\gamma_3$); $S_6$ and $S_7$ are noisy copies of $S_5$ with coupling strength $\gamma_2=0.9$; $S_{8}$ is defined via a noisy OR gate from $S_6$ and $S_7$, while $S_{2}$ is defined via a noisy OR gate from $S_3$, $S_4$ and $S_5$ (noisy gates have coupling strength $\gamma_1=0.9$ and are defined according to the conditional probabilities given in Fig. \ref{sim_bin}a).

Fig. \ref{sim_bin}b reports, for a simulation with $N=1000$ observations, the matrix of the B-index values computed using the plug-in estimator; non-significant estimates of MI or cMI leading to $B=1$, $B=-1$ or $B$ = NaN were assessed using random shuffling surrogates, and the network was reconstructed pruning the links with non-significant MI and/or cMI. The B-index correctly identified the presence and nature of the interactions imposed between pairs of variables, recovering for instance the existing isolated link ($B(S_9;S_{10})=0$, white) and the expected synergistic (e.g., $B(S_2;S_3)<0$, $B(S_2;S_4)<0$, blue) and redundant links (e.g., $B(S_5;S_6)>0$, $B(S_5;S_7)>0$, red). Moreover, it allowed to correctly prune out of the reconstructed network several links that would be detected by using only MI due to common drive or cascade effects (e.g., $\mathcal{B}(S_2;S_6)=1$, $B(S_5;S_8)=1$; dark red), or by using only cMI due to common target effects (e.g., $B(S_3;S_5)=-1$; dark blue). However, a false positive link was detected between the nodes $S_6$ and $S_7$: in spite of the lack of a direct connection, the two nodes result as connected because they have both a common drive ($S_6 \leftarrow S_5 \rightarrow S_7$) and a common target ($S_6 \rightarrow S_8 \leftarrow S_7$), inducing significant values for both MI and cMI and thus determining $-1<B(S_6;S_7)<1$.
This highlights a limitation of the B-index, which cannot guarantee the elimination of all spurious links.

To quantify the ability to correctly identify the presence and absence of links in the reconstructed network, we assessed the sensitivity and specificity of the B-index at varying the size $N$ of the data collected. Sensitivity and specificity, which measure respectively the impact of false negatives and false positives, were assessed over 100 simulation runs. The results in Fig. \ref{sim_bin}c highlight that the sensitivity increased markedly with the data length, approaching the expected value of $100 \%$ when $N=1000$.
On the contrary, the specificity showed less dependence on $N$, as it remained consistently high exhibiting minimal variations around the expected value of 97$\%$.

\subsubsection{Gaussian variables}

\begin{figure*}[btp]
\centering
\includegraphics[scale=1.15] {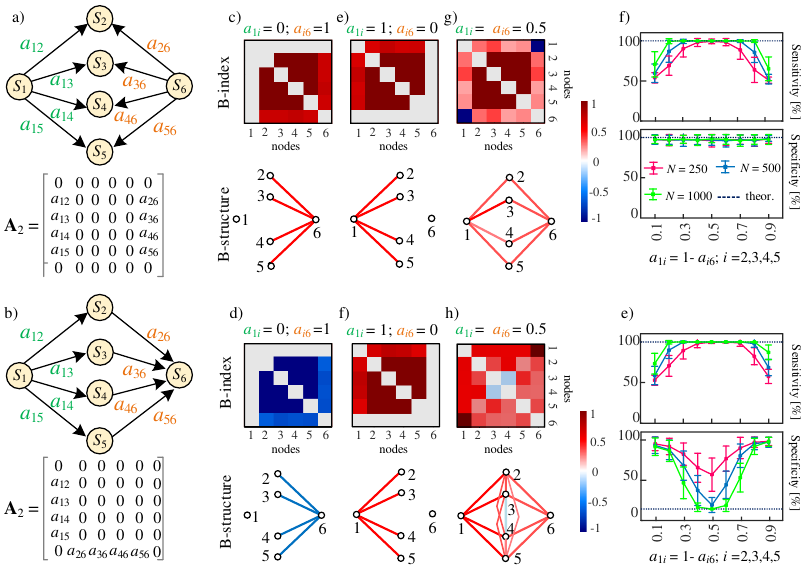}
\caption{Experimental analysis of a simulated dynamic system with six
units mapped by Gaussian processes connected via the the diagrams and VAR coefficients shown in (a) and (b).
The B-index rate matrix computed from multivariate time series ($N=1000$) and the corresponding reconstructed structures are investigated for peculiar parameter setting realizing: single star structures where the hub sends information to four leaves, observing accurate reconstruction of redundant links (c, e, f); a single star structure where the hub receives information from the leaves, observing accurate reconstruction of synergistic links (d); two competing star structures where both hubs send information to the same leaves, observing accurate reconstruction of redundant links (g); two interacting star structures with information propagated from one hub to the other via mediation of the leaves, observing reconstruction of redundant structures with false positive links among the leaves (h). 
The performance of network reconstruction, assessed over 100 simulation runs at varying the series length $N$ shows that sensitivity depends on $N$ and on link strength, while specificity is affected by simultaneous common drive and common target effects (i, j).}
\label{sim_stars}
\end{figure*}

The second simulation reproduces a network comprising $M=6$ nodes, whose activity is mapped by a multivariate random process defined by the VAR model (\ref{fullVAR}) implemented with $\mathbf{\Sigma}_U=\textbf{I}$, $p=2$, and time lagged coefficients reported in Fig. \ref{sim_stars}a,b. These settings produce two different connectivity structures, both characterized by two highly connected nodes (hubs, processes $S_1$ and $S_6$) interacting with four nodes with lower connectivity degree (leaves, processes $S_2 - S_5$). The hub $S_1$ acts as a source sending information to the leaves; the hub $S_6$ acts either as a source or as a sink, respectively sending information to the leaves which behave as receivers (Fig. \ref{sim_stars}a), or receiving information from the leaves which behave as mediators (Fig. \ref{sim_stars}b).

Fig. \ref{sim_stars}c,d reports, for different process realizations obtained as multivariate time series with $N=1000$ samples, the matrix of the B-index rates computed through linear parametric estimator, where non-significant values of the estimated MIR and cMIR leading to $B=1$, $B=-1$ or $B$ = NaN were assessed by using iAAFT surrogates, and the reconstructed network structure obtained by pruning the links with non-significant  MIR and/or cMIR.
The analyzed realizations reproduce the typical structures of hub-leaves interactions which are described and analyzed in the following. 
A single star structure (right-positioned) with the hub $S_6$ connected to the leaves $S_2 - S_5$ is obtained when $a_{1i}=0, i=2,\ldots,5$; the hub acts as a source in Fig. \ref{sim_stars}c and as a sink in Fig. \ref{sim_stars}d. These two configurations are correctly detected in terms of structure ($S_1$ is isolated and $S_6$ is connected to all leaves), and are differentiated by the clear redundancy (red) and synergy (blue) evidenced by the B-index rate in the two cases.
The configuration with hub sending information to the leaves, which results as fully redundant due to the dominance of common drive effects, is reproduced also when $a_{i6}=0$ ($i=2,\ldots,5$, left-positioned star), as shown in Fig. \ref{sim_stars}e,f where $S_1$ is the hub and $S_6$ is isolated.
The most rich configurations are those obtained setting nonzero values for all thee coefficients defining the structures in Fig. \ref{sim_stars}a,b. Specifically, when we set $a_{1i}=a_{i6}=0.5$ in Fig. \ref{sim_stars}a, a  configuration with two competing star structures where two hubs send information to the same leaves is obtained; in this case, all links are redundant due to the dominance of common-drive effects (except the link between the two hubs which is fully synergistic due to common target effects towards all leaves), and the network is reconstructed without errors (Fig. \ref{sim_stars}g).
On the other hand, setting $a_{1i}=a_{i6}=0.5$ in Fig. \ref{sim_stars}b, we obtain a configuration with propagation between two stars where one hub sends information to the other through mediation of the leaves; in this case, the links are again dominantly redundant due to the abundance of common drive and cascade effects, but the network reconstruction suffers from false-positive detections of links between the leaves caused by the simultaneous presence of common drive and common target effects determining significant MIR and cMIR.

Finally, the accuracy of network reconstruction was investigated for different time series length $N$ at increasing
%, in the network structures of Fig. \ref{sim_stars}a,b,
the weight of the connections from the hub $S_1$ to the leaves $S_2 - S_5$ while simultaneously decreasing the weight of the connections between the leaves and the hub $S_6$  ($a_{1i} \in [0.1, 0.9], a_{i6}=1-a_{1i}$), so as to gradually move from single-star to two-star structures and back. The results reported in Fig. \ref{sim_stars}i,j show that the sensitivity depends on the data length and on the link strength, reaching the expected value of $100 \%$  when the links are balanced in strength. 
The specificity was very high and substantially unaffected by the data length in the case of competing stars (Fig. \ref{sim_stars}i), while it was higher than expected for unbalanced link strength and approaching the expected value of $14.3 \%$ only for balanced link strength and high $N$ (Fig. \ref{sim_stars}j). These results confirm the good performance of the B-index in reconstructing the statistical structure underlying directed networks, with the limitations related to its inability to resolve conditions of contemporaneous common drive and common target effects impinging on a link.

\section {Applications to Cardiovascular Networks}

This section reports the application of the proposed methodology to physiological networks probed measuring simultaneously the spontaneous variability of several \textcolor {black} {cardiovascular (cardiac and hemodynamic)} and respiratory variables during different physiological states.
Specifically, in the first application we perform a static analysis of the discrete random variables representative of the activity of the \textcolor {black} {cardio}, vascular and respiratory systems, whose observations describe respectively the acceleration/deceleration of the heart rate, \textcolor {black} {the variation of blood pressure as the basic parameter of hemodynamics}  and the phase of the respiratory cycle (inspiration/expiration).
In the second application we perform a dynamic analysis of five continuous random processes descriptive of the beat-to-beat variability of heart period, systolic and diastolic pressure, cardiac output and peripheral resistance.
Both analyses are performed considering short-term cardiovascular variability in groups of healthy subjects monitored in resting state and during postural stress \cite{faes2017information,krohova2020vascular}.

\subsection {Subjects and Experimental Protocols}
Analyses were performed on two groups of young healthy volunteers, both recruited at the Jessenius Faculty of Medicine, Comenius University, Martin, Slovakia, where ethics approval and informed consent were obtained.
%The two studies involved respectively 61 volunteers (37 females, $17.5 \pm 2.4$ years), and 39 volunteers (22 women, age $19.4 \pm 2.3$ years).
For both studies, subjects were monitored following the same experimental protocol consisting of recording cardiovascular signals for 15 minutes in the resting supine position (REST), followed by further 8 minutes of recordings in the upright position (TILT) reached after tilting the subjects to 45 degrees the motor-driven bed table to evoke mild orthostatic stress. Details about the protocol can be found in Refs. \cite{faes2017information, javorka2017basic} and \cite{krohova2020vascular}.
 
The first study involved 61 volunteers (37 females, $17.5 \pm 2.4$ years), in whom the electrocardiogram (ECG), 
%(CardioFax ECG-9620, NihonKohden, Tokyo, Japan)
the continuous finger arterial blood pressure (ABP) collected noninvasively by the photoplethysmographic volume-clamp method, %(Finometer Pro, FMS, Netherlands)
and the respiratory signal (RESP) obtained via inductive plethysmography using thoracic and abdominal belts
%(RespiTrace 200, NIMS, Miami Beach, FL, USA)
were recorded simultaneously.
The second study involved 39 volunteers (22 women, age $19.4 \pm 2.3$ years), in whom the ECG and ABP signals were measured as before simultaneously with the impedance cardiography signal.
%(ICG, CardioScreen 2000, Medis, Germany).
In both cases, signals were digitized with 1 KHz sampling rate.

\subsection {Data Analysis}

In the first application, the beat-to-beat time series of the heart period ($HP$), systolic pressure ($SP$) and respiratory amplitude ($RA$) were measured from the ECG, ABP and RESP signals acquired for each subject and condition respectively as the sequences of the temporal distances between consecutive R peaks, the maximum ABP values within each detected R-R interval, and the RESP values taken at the onset of each detected R-R interval, according to a well-established measurement convention \cite{faes2017information,javorka2017basic}. For each of these sequences measured synchronously over $N$ consecutive heartbeats, realizations of the three discrete random variables considered for the static analysis were obtained as follows ($n=2,\ldots,N-1$). 
The heart rate variation $HV$ was obtained observing the sign of the difference between the duration of two consecutive heart periods: $HV_n=0$ if $HP_{n+1} \leq HP_{n}$ (acceleration), $HV_n=1$ if $HP_{n+1}>HP_{n}$ (deceleration). The systolic pressure variation $SV$ was obtained similarly: $SV_n=0$ if $SP_{n} \leq SP_{n-1}$, $SV_n=1$ if $SP_{n}>SP_{n-1}$. The respiration phase $RP_n$ was obtained by discretizing the respiration amplitude $RA_{n+1}$; specifically $RP_n = \overline{RA}_{n+1}$, where $\overline{RA}_{i}$ = 0 if $RA_i \leq RA_{i+1}$ and  $\overline{RA}_{i}$ = 1  if $RA_i > RA_{i+1}$.
%\textcolor {blue} {$HV_n=0$ if $HP_{n+2} \leq HP_{n+1}$ (acceleration), $HV_n=1$ if $HP_{n+2}>HP_{n+1}$ (deceleration), where $n=1, \dots, N=300-2$. The systolic pressure variation $SV$ was obtained similarly: $SV_n=0$ if $SP_{n+1} \leq SP_{n}$, $SV_n=1$ if $SP_{n+1}>SP_{n}$. The respiration phase $RP_n$ was obtained by discretizing the respiration amplitude $RA_{n+2}$; specifically $RP_n = \overline{RA}_{n+2}$, where $\overline{RA}_{i}$ = 0 if $RA_i \leq RA_{i+1}$ and  $\overline{RA}_{i}$ = 1  if $RA_i > RA_{i+1}$.}

\begin{figure} [tbp]
\centering
\includegraphics[scale=0.78] {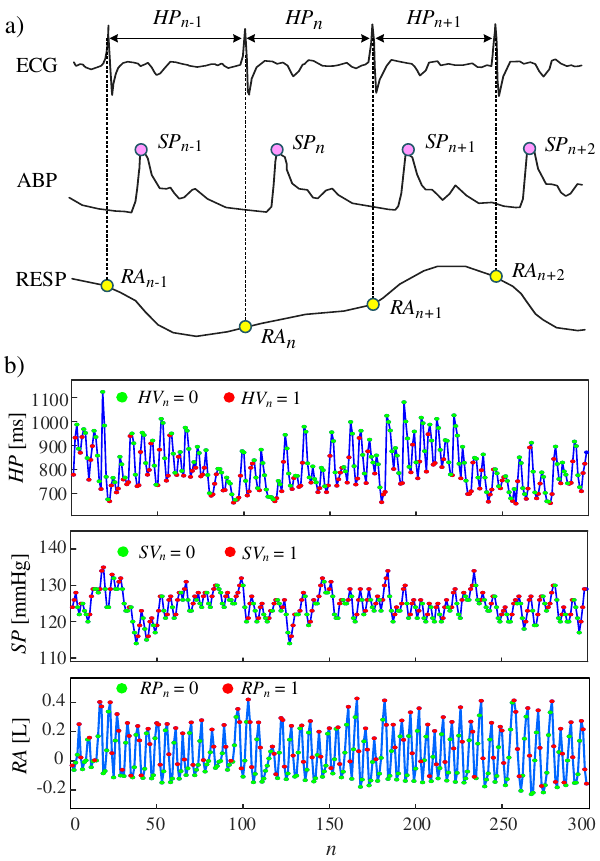}
\caption{a) Bio-signal processing and measurement convention for the extraction of the time series used in the static analysis of cardiovascular and respiratory interactions. (a) The ECG, ABP and RESP signals are used to measure the beat-to-beat parameters corresponding to heart period $HP$, systolic pressure $SP$ and respiration amplitude $RA$. (b) These parameters are then used to build observations of the three discrete random variables analyzed in the static analysis, i.e. the heart rate variation $HV$, the systolic \textcolor {black} {pressure variation $SV$}, and the respiratory phase $RP$.}
\label{app1_sig}
\end{figure}

%The systolic  level $SL$ was obtained comparing the systolic pressure samples with two thresholds $SP_{33}$ and $SP_{66}$ set at the $33^\mathrm{th}$ and $66^\mathrm{th}$ percentiles of the distribution of the systolic pressure values: $SL_n=0$ if $SP_{n-1}<SP_{33}$ (low pressure), $SL_n=1$ if $SP_{33} \leq SP_{n-1} \leq SP_{66}$ (medium pressure), and $SL_n=2$ if $SP_{n-1} > SP_{66}$ (high pressure). 
%The respiration phase $RP$ was determined comparing two consecutive respiration amplitude values: $RP_n=0$ if $RA_n \leq RA_{n+1}$ (inspiration), $RP_n=1$ if $RA_n > RA_{n+1}$ (expiration).} 
The adopted measurement convention and the rules set to measure the three discrete variables are illustrated in Fig. \ref{app1_sig}a.
For each subject, realizations of the variables $HV$, \textcolor {black} {$SV$} and $RP$ were obtained as described above from sequences of $N=300$ consecutive $HP$, $SP$ and $RA$ values selected during stationary epochs of the REST and TILT phases (an example is reported  Fig. \ref{app1_sig}b). Then, the static analysis of the network with $M=3$ nodes was performed estimating MI and cMI through the plug-in method, assessing their significance through the use of 100 random shuffling surrogates (significance $\alpha=0.05$), and finally computing the B-index from the thresholded MI and cMI.

\begin{figure} [tbp]
\centering
\includegraphics[scale=0.8] {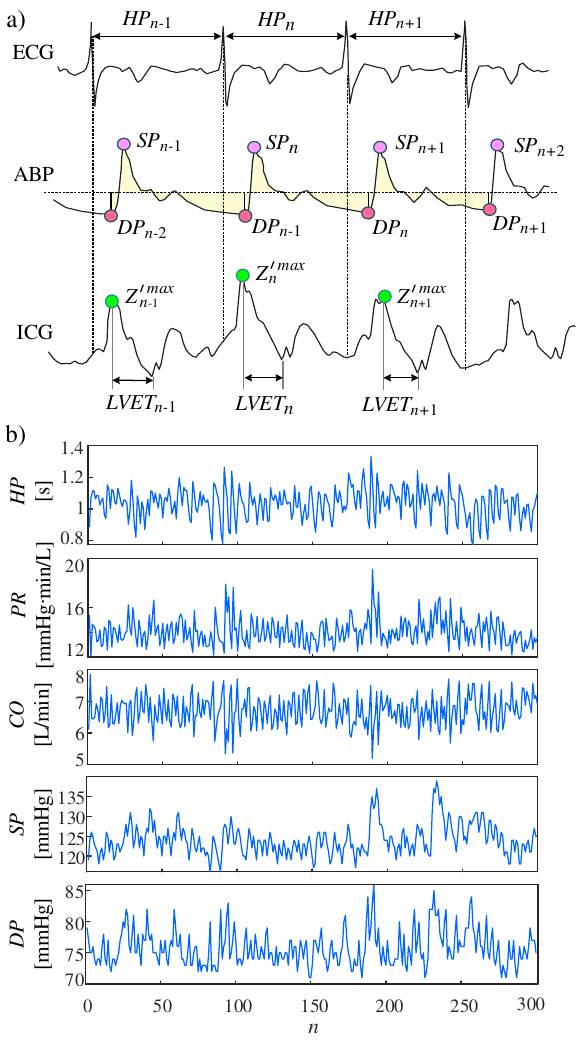}
\caption{Bio-signal processing and measurement convention for the extraction of the time series used in the dynamic analysis of cardiovascular networks. The signals and parameters used to build the five analyzed time series are shown in (a): the ECG and ABP signals are used to measure the heart period $HP$, the systolic pressure $SP$ and the diastolic pressure $DP$; the ICG signal is differentiated to measure the local maxima of the impedance variations $Z'^{max}$ and the left ventricular ejection time $LVET$ from which the cardiac output $CO$ is derived; and the ICG and ABP signals are combined to measure $CO$ and the mean arterial pressure (yellow shades) from which the peripheral resistance $PR$ is derived.
Representative time series measured for a subject monitored in the resting supine position are shown in (b).}
\label{app2_sig}
\end{figure}

In the second application, the analyzed beat-to-beat time series were the $HP$, $SP$, diastolic pressure ($DP$), cardiac output ($CO$), and peripheral vascular resistance ($PR$) obtained from the ECG, ABP and ICG signals as follows: $HP_n$ is the duration of the current R-R interval; $SP_n$ is the maximum ABP value measured within $HP_n$; $DP_n$ is the minimum ABP value measured between the occurrence times of $SP_n$ and $SP_{n+1}$; $CO_n= 60 \cdot \frac{SV_n}{HP_{n-1}}$, where the stroke volume is computed as $SV_n=\beta \cdot Z'^{max}_n \cdot LVET_n $, being  $LVET_n$ the left ventricular ejection time, $Z'^{max}_n$ the maximum of the time-derivative of the impedance signal taken within $HP_n$, and $\beta$ a correcting factor accounting for thorax volume and base impedance; and $PR_n=\frac{MAP_n}{CO_n}$, where $MAP_n$ is the mean ABP measured between the occurrence times of $DP_{n-1}$ and $DP_{n}$. 
%the sequences of the R-R intervals, the maximum ABP values within each detected R-R interval, the minimum ABP values detected between two consecutive systolic values, the ratio between the current stroke volume and the previous R-R interval ($CO_n=60 \cdot SV_n$ where the stroke volume $SV$ was computed at each heartbeat from the time-derivative of the impedance signal taking its local maxima and multiplying them by the left ventricular ejection time and by a corrective factor according to the Bernstein and Sramek formula \textcolor{red}{[40Jana]}), and the ratio between the current mean blood pressure and cardiac output.
This measurement convention is typically adopted in computational physiology \cite{bernstein1986continuous,javorka2017basic} and is illustrated in Fig. \ref{app2_sig}a.
For each subject and condition, stationary realizations of $N=300$ points of the multivariate process $\{HP, SP, DP, CO, PR \}$ were obtained as described above (an example is reported in  Fig. \ref{app2_sig}b). Then, the dynamic analysis of the network with $M=5$ nodes was performed estimating MIR and cMIR through the regression-based method, implemented setting the VAR model order through the Akaike Information Criterion \cite{akaike1974new} and using $q=20$ lags to identify the restricted VAR models, assessing the significance of MIR and cMIR through the use of 100 iAAFT surrogates (significance $\alpha=0.05$), and finally computing the B-index rate from the thresholded MIR and cMIR.

\subsection {Results and Discussion}

In the first application the processing of the $HP$, $RA$ and $SP$ time series described in Sect. IV.B led us to investigate, through a simple static analysis of the discrete random variables $HV$, $SV$ and $RP$, three types of interaction related to known physiological mechanisms, i.e. the arterial baroreflex (link $HV-SV$), the respiratory sinus arrhythmia (link $HV-RP$) and the mechanical effects of respiration on arterial pressure (link $RP-SV$) \cite{malpas2002neural,cohen2002short}.

%\bigskip

\begin{figure} [tbp]
\centering
\includegraphics[scale=0.74] {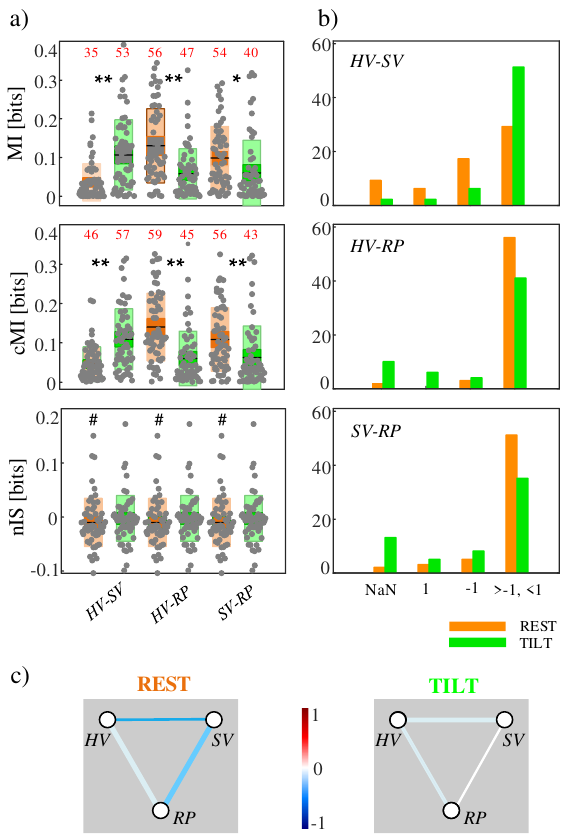}
\caption{Results of the static analysis of cardiovascular and respiratory interactions. (a) Distribution across 61 subjects of the mutual information (MI), conditional MI (cMI) and net information shared (nIS = MI$-$cMI) between the discrete variables representing heart rate variation ($HV$), systolic pressure \textcolor {black} {variation ($SV$)} and respiratory phase ($RP$) measured in the supine position (REST) and during postural stress (TILT); \textcolor {black}{asterisk and hash symbols denote statistically significant differences between REST and TILT, and between MI and cMI  respectively (*,\#: $p<$ 0.05, **: $p<$ 0.005 paired Wilcoxon test)}. (b) Number of subjects for which both MI and cMI were not statistically significant ($B$-index = NaN), only MI was significant ($B=1$), only cMI was significant ($B = -1$), or both MI and cMI were significant ($|B|<1$) in the two conditions; the significance was assessed using random shuffling surrogates. (c) Reconstructed network structure where the link thickness is associated to the number of subjects with significant MI and cMI and the link color maps the average B-index.}
\label{app1_res}
\end{figure}

The analysis of the information shared between these variables shows that the dominant interactions are identified by the links $HV-RP$ and $SV-RP$ at rest, and by the link $HV-SV$ during tilt, as documented by the high and statistically significant values of both MI and cMI (Fig. \ref{app1_res}a,b). The prevalence of the interactions involving the respiratory activity (variable $RP$) in the resting condition reflect the well known effects of respiration on the variability of the heart rate and of respiration. These effects appear weaker during the postural challenge, likely reflecting the dampening of respiratory sinus arrhythmia with postural stress \cite{faes2011information}, though remaining significant in more than half of the subjects.
On the other hand, the stronger interaction $HV-SV$ during postural stress can be ascribed to the enhancement of the baroreflex activity associated with sympathetic activation, which determines acceleration (deceleration) of the heart rate in correspondence with  decrease (increase) of the systolic pressure levels \cite{cooke1999human,faes2013mechanisms}.

%\bigskip

\bigskip

The analyzed interactions were detected as significantly synergistic during rest, and exhibiting a net balance between synergy and redundancy during tilt (Fig. \ref{app1_res}a,c; note that the nIS values are the same for all links in this application with three variables analyzed). 
The small but statistically significant prevalence of synergy in the resting state documents the existence of a dominant common target effect (see also the simulated systems of Fig. 2); in this application, the common target situation arises likely by the simultaneous presence of baroreflex effects from $SV \rightarrow HV$ and of direct respiratory sinus arrhythmia $RP \rightarrow HV$ \cite{cooke1999human}.
The balance between redundancy and synergy induced by the postural stress might indicate a balancing of the two pathways whereby respiration affects the heart rate, resulting from a wakening of the direct respiratory sinus arrhythmia (pathway $RP \rightarrow HV$) and a strengthening of the baroreflex-mediated respiratory sinus arrhythmia (pathway $RP \rightarrow  SL \rightarrow  HV$).

\begin{figure} [tbp]
\centering
\includegraphics[scale=0.73] {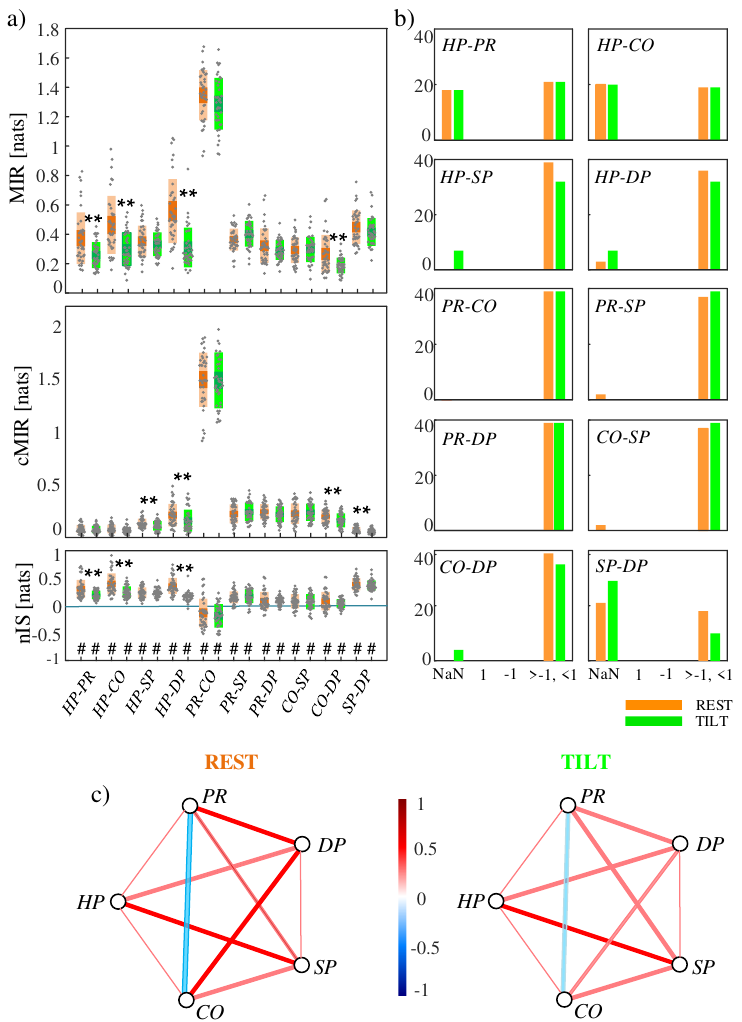}
\caption{Results of the dynamic analysis of cardiovascular networks. (a) Distribution across 39 subjects of the values of mutual information rate (MIR), conditional MIR (cMIR) and net information shared (nIS = MIR$-$cMIR) between the continuous processes representing the variability of heart period ($HP$), systolic pressure ($SP$), diastolic pressure ($DP$), cardiac output ($CO$) and peripheral resistance (PR) measured in the supine position (REST) and during postural stress (TILT); black and blue hash symbols denote statistically significant differences between REST and TILT, and between MI and cMI (p$<$0.05, paired Wilcoxon test). (b) Number of subjects for which both MIR and cMIR were not statistically significant ($B$-index=NaN), only MIR was significant ($B=1$), only cMIR was significant ($B=-1$), or both MIR and cMIR were significant ($|B|<1$) in the two conditions; the significance was assessed using random iAAFT surrogates. (c) Reconstructed network structure where the link thickness is associated to the number of subjects with significant MIR and cMIR and the link color maps the average B-index rate.}
\label{app2_res}
\end{figure}

In the second application, the dynamic formulation of the proposed framework was exploited to explore in detail the interactions among several cardiovascular variables related to cardiac rhythm and \textcolor {black} {hemodynamics, including the regulation of arterial pressure}. The results in Fig. \ref{app2_res} indicate that four of the nodes of this network, i.e. those mapping the variability of $PR$, $CO$, $DP$ and $SP$, form a fully connected subnetwork which is stable in the two analyzed conditions. This finding is documented by the statistically significant values of both MIR and cMIR, as well as of their balance, consistently observed at rest and during tilt for the links $PR-CO$, $PR-SP$, $PR-DP$, $CO-SP$ and $CO-DP$ (Fig. \ref{app2_res}a,b).
In particular, the link between $PR$ and $CO$ is very strong and exhibits the highest values of both MIR and cMIR, as a consequence of the inverse relation existing between the two processes. Notably, for this link cMIR is consistently higher than MIR, resulting in significantly negative values of the interaction information rate; this statistically significant synergy reveals the existence of a common target relation from $PR$ and $CO$ towards other connected variables in the network, likely $SP$ and especially $DP$.

Physiologically, this relation might be associated to the effect \textcolor {black}
{of the variations in vasomotion characterized by the vascular resistance dynamics on the blood pressure and arterial compliance} \cite{bank1995direct, nardone2018evidence}. Interestingly, when $PR$ or $CO$ are analyzed together with one of the vascular processes (either $SP$ or $DP$), the link becomes significantly redundant; this suggests that $HP$ plays a role through its tight relation with $SP$ and $DP$ reflecting known physiological effects like the baroreflex and the cardiac run-off \cite{javorka2017basic}.
The absence of significant changes of nIS and B-index with tilt for these links suggests that the underlying mechanisms are not modulated by sympathetic activation \textcolor {black} {or parasympathetic inhibiton associated with orthostasis}.

Different behaviors were observed when the links including heart rate variability were considered. The links between $HP$ and $PR$ or $CO$ were detected in only half of the subjects in both conditions (Fig. \ref{app2_res}b), and were characterized by redundancy (MIR $>$ cMIR) which decreased moving from rest to tilt due to decreased MIR and unaltered cMIR.
The links between $HP$ and $SP$ or $DP$ were detected in the majority of the subjects in both conditions, and were again characterized by redundancy with a decrease from rest to tilt in the case of $HP-DP$, and stable net redundancy in the case of $HP-SP$.
These results indicate that the sympathetic activation evoked by tilt tends to make  the interactions between $HP$ and the other cardiovascular processes less redundant.
The known activation of the baroreflex mechanism with tilt was not evidenced by the link $HP-SP$, possibly because the symmetric measures used here (MIR, cMIR) account for both feedback and feedforward cardiovascular interactions \cite{faes2013mechanisms}.
\textcolor {black} {The interaction $SP-DP$, which physiologically may be related to the Frank-Starling effect \cite{javorka2017basic}, was significantly found as redundant in about half of the population, with a tendency to decrease during tilt}.
We note that, since the respiratory activity plays a main role in cardiovascular control \cite{cohen2002short}, these results should be complemented including respiration in the network, or performing spectral analysis to focus on low-frequency oscillations.

\section{Conclusions and Perspectives}

The framework proposed in this work for the analysis of physiological networks is designed to evaluate how two nodes are functionally connected and interact with the rest of the network. This approach makes the proposed measures of synergy/redundancy balance fundamentally different from the existing ones. In fact, while HOI measures based on partial information decomposition \cite{williams2010nonnegative} concentrate on one network node and relate its activity to that of two or more other nodes, and the recently proposed O-information measures \cite{rosas2019quantifying,faes2022new} concentrate on the whole network analyzed collectively, the nIS measure puts the focus on each specific link, thus allowing to represent high-order effects as networks.
Moreover, the normalization leading to the B-index and the associated analysis for statistical significance allow to prune the indirect links determined by cascade/common drive or common target relations, thus making it possible to exploit HOIs for inferring the structure of the analyzed functional network.

The validation of the framework on theoretical and numerical simulated networks suggested its ability to catch the balance between synergy and redundancy effects characterizing the interactions between two nodes and the rest of the network, as well as to reconstruct the network structure. While the characterization of static and dynamic networks was performed implementing respectively model-free and model-based estimators, the two approaches can be applied to both types of networks. Nevertheless, the estimation performance degrades at decreasing the data size and increasing the network size, especially for the model-free estimator, due to known statistical effects and to the curse of dimensionality. Moreover, even in optimal conditions, structural dependencies involving both common drive and common target mechanisms cannot be handled by the B-index and result in false positive detections. This constitutes a limitation that could be solved theoretically by accounting for all possible subset of nodes while testing for conditional dependencies, but this becomes intractable in practice when increasing the network size. These aspects should be considered in extensions of the framework incorporating refined entropy estimators \cite{kraskov2004estimating} and implementing criteria for dimensionality reduction \cite{faes2011information}.

The application to physiological networks showed how the proposed framework can elicit mechanisms of cardiovascular regulation investigated at rest and in response to postural stress. We purposely implemented both static and dynamic analyses to illustrate the flexibility of our framework: while in principle a dynamic analysis of random processes is more complete and allows a more fine-grained characterization of diverse interactions, the static analysis allows to track important physiological mechanisms also starting from coarse-grained parameters which can be obtained faster and more robustly.
Thus, depending on the applicative context, the use of the proposed static and dynamic measures of synergy/redundancy balance and link strength can favor a deeper investigation of physiology as well as the discovery of new clinical markers. These measures can also be exploited to complement existing analyses for empowering the automatic classification of \textcolor {black} {pathophysiological} states.

\bibliography{BindexREF}
\end{document}